# Visual Milestone Planning in a hybrid development context

Eduardo Miranda [0000-0001-8195-7506]

Carnegie Mellon University, Pittsburgh PA 15213, USA

`mirandae @ andrew.cmu.edu`

**Abstract.** This paper explains the Visual Milestone Planning (VMP) method using an agile vocabulary to facilitate its adoption by agile practitioners as a front end for a hybrid development process. VMP is a visual and collaborative planning approach which promotes a shared understanding of the work approach and commitment through the direct manipulation by team members of the reified planning constructs involved in the development of the plan. Once the product backlog has been established and relevant milestones identified, a novel construct called the milestone planning matrix is used to document the allocation of product backlog items to milestones. The milestones due dates are later determined by grouping sticky notes representing the work to be performed into timeboxes called work packages and accommodating them on a resource and time scaled scheduling canvas very much as it would be done in a Tetris game.

## 1      Introduction

Prominent agile authors have long advocated the need for a project to have an artifact to guide the work of a team through it. Cohn [1], for example, suggests the use of a release plan, without which "teams move endlessly from one iteration to the next"; Cockburn [2], adds "a coarse-grained project plan, possibly created from a project map or a set of stories and releases to make sure the project is delivering suitable business value for suitable expense in a suitable time period"; Highsmith [3], postulates a Speculate Phase, in which "a capability and/or feature-based release plan to deliver on the vision is developed as well as a wave (or milestone) plan spanning several iterations used as major synchronization and integration points"; and Brechner [4] writes "it's important to have a vision and a plan for achieving your project goals. Even crowdsourcing projects need an organizing principle and structure to allow everyone to contribute toward the shared outcome".

Although these authors discuss the characteristics these plans must show, e.g. that the plan must be formulated, not in terms of the tasks to be performed but rather in terms of the outcomes the project must deliver and how they must be elaborated, e.g.

collectively by the team and not by a solitary project manager who later hands it down to it for execution, they do not provide a method for doing it.

This approach to planning was first proposed by Andersen [5] who called it Milestone Planning. In his seminal article, Andersen defines a milestone "as result to be achieved … a description of a condition or a state that the project should reach by a certain point in time. A milestone describes what is to be fulfilled, but not the method to fulfil it". In a subsequent work [6], he describes milestone planning as an activity to be performed by the group stating "We strongly emphasize the motivational and inspirational aspects of planning. They are often neglected in practice so that planning becomes a tedious chore carried out on the project manager's desk or PC. This results in a lack of ownership of the plan by the parties involved in the project and consequently the plan is never actively used. This is one reason for the failure of so many projects", but also like in the case of the agile authors, he failed to provide a method to construct such plans. The gap was addressed by Miranda [7] who proposed a participatory and visual approach to construct milestone plans he called the Visual Milestone Planning (VMP).

By their own nature, good milestone plans are robust, comprehensive, easy to understand, and confer great flexibility in terms of how to achieve the milestones, all of which makes them a good fit to organize agile endeavors.

This paper casts the VMP approach in the context of planning a project which will be executed using an agile approach like Scrum. Creating what is commonly defined as a hybrid approach [8]. The execution aspects of the project are only cursory covered due to page count limits and the interested reader is directed to [9] and [10] for an in-depth treatment.

The rest of the paper is organized as follows: Section 2 describes milestone plans, Section 3, briefly describe the execution of the plan, Section 4 explains the proposed method, Section 5 provides a detailed example that illustrates the use of the method and serves as validation and Section 6 provides a summary and reference to the initial evaluation of the method.

## 2      Milestone plans

Figure 1 shows a conventional milestone plan. As can be observed, the plan is concise, typically confined to a size which will allow it to be grasped at once and written using a vocabulary stakeholders can understand. The plan comprises the sequence of states the project will go through, from its inception to its successful conclusion, and not the activities the team needs to perform to achieve those states which will be proposed as work progresses. For example, the "Design concept approved" milestone defines a state where the project team has presented an idea that satisfies the needs of the sponsor, and she has acquiesced to it. The plan does not estipulate how the team will get there. Will they show wireframe diagrams to her? Develop high fidelity prototypes? Make a PowerPoint presentation? These issues will certainly have to be addressed by the team, for example during sprint planning, but they have no place in the milestone plan.

The focus on states rather than on activities results in a more robust plan since, independent of what tasks need to be performed to get there, when and by whom, it is unlikely, the project sponsors' desire to approve the design concept before it is implemented, will change.

The dependencies between milestones are typically "Finish to Finish" relations, meaning that if "Milestone B" depends on "Milestone A", "Milestone B" cannot be completed until "Milestone A" has been completed. Finish to Finish relations are easy to spot and provide flexibility as to when the activities leading to the realization of the milestone could start.

Milestones could be hard or soft. Hard milestones are milestones, that if not accomplished by a set date, lose all or most of its value or results in severe penalties. The date a government resolution which the system under development is supposed to address goes into effect and the start of the holidays shopping season are examples of hard milestones a project might encounter. Soft milestones, on the other hand, have completion dates that result from the planning process. They might be associated with penalties or other liabilities after a statement of work is agreed, but in principle are discretionary.

| State[1] | Planned date | Forecasted date | Team | Client | Description | Responsible |
|---|---|---|---|---|---|---|
| | Oct 1st | | | | Project kick-off | |
| | Oct 20th | | | | UX concept approved | |
| | Nov 1st | | | | Cloud infrastructure selected | |
| | Nov 10th | | | | Ux design completed | |
| | Jan 10th | | | | Cloud infrastructure available | |
| | Jan 15th | | | | Release 1: CL, BD, AC, CO, Data Base | |
| | Jan 20th | | | | Beta testing launched | |
| | Feb 10th | | | | Release 2: CBL, RC, SM | |
| | Feb 25th | | | | Beta testing results reviewed | |
| | Mar 10th | | | | Release 3: Customer feedback + emergent features | |
| | Mar 15th | | | | Acceptance testing procedure approved | |
| | Apr 1st | | | | Acceptance test completed | |
| | May 1st | | | | System deployed | |
| | May 15th | | | | Customer sign-off | |
| | May 15th | | | | Project closed | |

[1]State: C – Completed, P – In progress, Blank – Not started

◯ - Hard milestone   ⬡ - Soft milestone

**Fig. 1.** A typical milestone plan showing due dates, responsibilities and milestones' descriptions

Miranda [10], supplemented Andersen's original technique, with the concept of "Work Packages Schedule" (see Figure 2). A Work Packages Schedule consists of a number of timeboxes within which all the work associated with a given milestone, called its work package, will have to be executed for the plan to hold. The purpose of the Work Packages Schedule is twofold: 1) to determine the earliest day by which a milestone can be completed in the context of other project work that might need to be performed, and 2) to drive the scheduling of work during project execution.

Within the constraints imposed by the hard milestones' due dates and the dependencies identified in the plan, the Work Packages Schedule will be decided by the team according to its technical, business and staffing strategies, such as: this needs to be done before that, do as much work as possible at the onset, start slow to minimize risk and then aggressively ramp up, maintain a constant workforce, work must be completed within six months, do not use more than five people, and so on. In constructing it, we will assume, the distribution of competencies in the team matches the work's needs. This is a sensible assumption in an agile context which assumes either generalists or balanced, cross-functional, teams. In cases where this assumption would not hold, it would be possible to break the resource dimension into competency lanes and assign the corresponding effort to each lane. The same approach could be used to scale up the method to be used in projects with multiple teams.

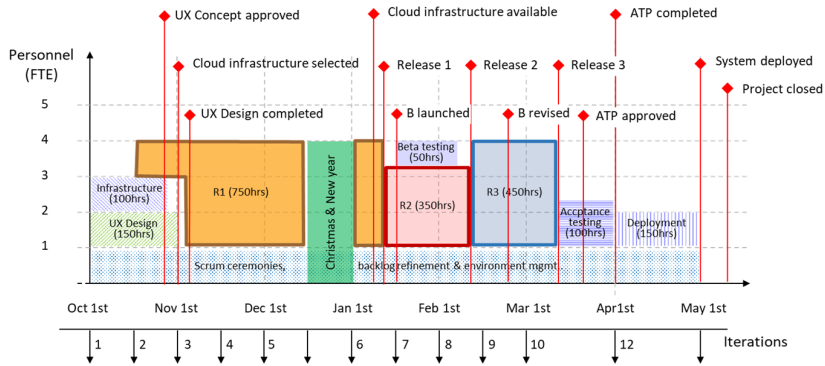

**Fig. 2.** Work Packages Schedule. This represents one possible arrangement of work packages corresponding to the milestone plan in Figure 1. Each shaded area corresponds to the work associated with the milestone immediately to its right. The resource-time frame enclosing the work package is its timebox. During iteration 1 and part of iteration 2, one team member will work on UX Design and another in the selection of infrastructure. Other team members can help as needed. From iteration 2 to 6, the team will mainly work on the items included in the first release; from iteration 7 to 9, the team will work on the features included in the second release and in Beta testing. Adapted from [7]

## 3    Executing the plan

In the course of the project, the team progressively refines items in the backlog to meet their definition of ready and decomposes them into the tasks necessary for its realization during the iteration/sprint planning meeting, as dictated by the timeboxes depicted by the Work Packages Schedule instead of the biweekly concerns of the product owner. As development progresses, the plan is updated to reflect new circumstances arising from the work completed or from changes in the project context, but since milestones are basically states or goals to be attained, and the plan does not specify when tasks must begin, how long they should take, nor who should perform them, it tends to be pretty stable.

## 4      The Visual Milestone Planning (VMP) Method

Figure 3 depicts the VMP Method. The first step in the process is to create a product backlog defining the project scope. Each backlog item in it will have associated the estimated effort required for its realization. These estimates will be later used to establish the timeboxes in which the anticipated work could be performed.

VMP proposes the adoption of a hierarchical product backlog, see Figure 4, which is an enumeration of all the outputs to be delivered to the customer to meet the project objectives: functionality, documentation, infrastructure, gadgets, and services, decomposed into smaller chunks, until they reach the user story level, which defines backlog items that can be completed in a single iteration. Technical stories are work the team needs to perform but unlike user stories do not have a direct user value counterpart. The hierarchical nature of the backlog facilitates the comprehension of the project's scope and supports the progressive refinement of the items identified, their estimation and the collective assignment to milestones. The structure proposed is compatible with the definition of backlog posited nowadays by most practitioners and tools, e.g. SAFe (epics, capabilities, features, stories), Jira (initiatives, epics, stories), Azure (epics, features, stories), Rubin (epics, themes, stories).

Since the backlog forms the basis for planning, it cannot be open ended in the sense of a backlog in a traditional agile project. Balancing predictability with progressive refinement requires the creation of planning aggregates for which a budgetary allowance will be made, in the understanding that when the time comes and they are defined, either we will circumscribe our level of ambition to the available budget, or the plan will need to be revisited.

A project whose final outcome is not well defined could be scoped in terms of learning activities such as running a design sprint and planning packages that have a definite purpose, e.g. Release 1, Release 2, etc., and budget, but whose exact content has not yet been decided.

The second step in the process is the definition of milestones. Milestones are chosen to signal the attainment of a major achievement, the delivery of key components or assemblies, the completion of important process steps or to mark a commitment made to the team, e.g., a customer makes proprietary equipment or technology required by the project available to the team. Notice that in the diagram there are arrows back and forth between steps 1 and 2. This is so because although the product backlog will normally inform the choice of milestones, sometimes the establishment of a particular milestone might result in the creation of new backlog items that must be incorporated to it or leads to its rearrangement.

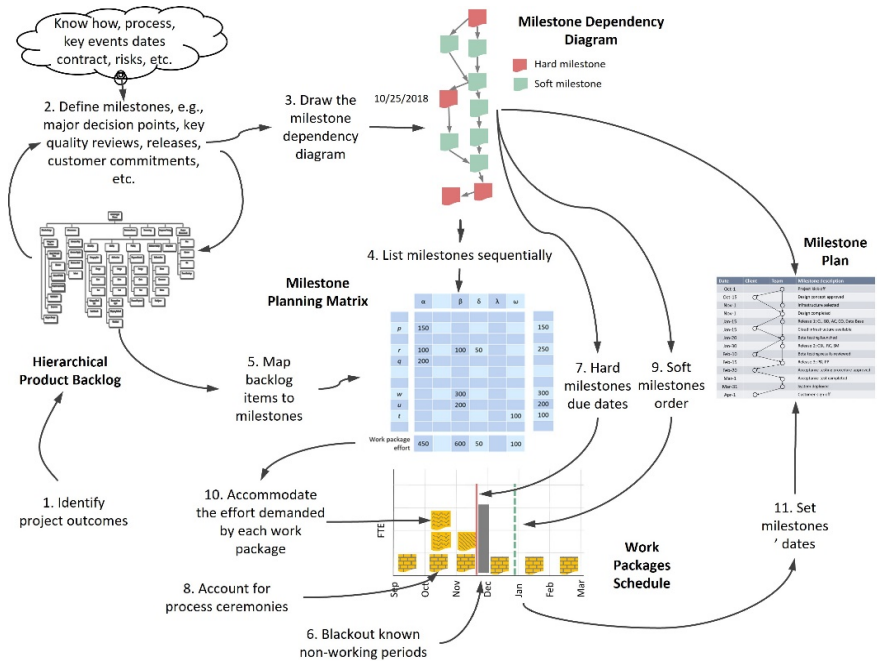

**Fig. 3.** Visual Milestone Planning Method

In step 3, we document the dependencies existing between milestones by means of the Milestone Dependency Diagram. Notice the diagram contains no dates, except for those associated with hard milestones. This is to permit the consideration of different staffing and timing strategies later in the process. The process for the creation of the Milestone Dependency Diagram starts with the writing of each identified milestone name on a sticky note1 and its quick ordering, according to their most obvious sequence of completion, followed by a discussion of specific dependencies, the connecting of the corresponding milestones, and, if necessary, the reordering of the original sequence. The dependencies between milestones are finish-to-finish, since these are more obvious than the most common finish-to-start and give a lot of flexibility on when to start working on a work package. A simple example of a finish-to-finish dependency is that between coding and testing. One could start writing test cases before coding begins, but one cannot finish it until the coding is done. In the fourth step, each cell in the header row of the Milestone Planning Matrix is populated with the name of the milestones.

---

[1] Although it is possible to do this using digital tools, it is recommended at least initially to do it this way because VMP promotes involvement and commitment, through the reification of the planning constructs: work packages, milestones and schedules employed in the planning process and their direct manipulation by the team members who collectively create the plan.

Although not strictly required by the process, listing them chronologically from left to right greatly contributes to the matrix readability and ease of work.

In step 5, backlog items are associated with the milestones they help realize via the body of the Milestone Planning Matrix. See Figure 5. The association is informed by the milestone definition, for example, a "Release 1" milestone would be associated with all backlog items included in the said release. The association is done by labeling a row in the planning matrix with the name of the top-most backlog item whose descendants all contribute to the same milestone and recording the effort required by it at the intersection of the row with the column corresponding to the milestone with which the item is being associated. A milestone can have multiple backlog items associated with it. In most cases, backlog items would be associated with a single milestone in its entirety; there are, however, a circumstances like planning packages or integrative efforts where it is convenient to allocate fractions of the total effort required by the item to multiple milestones.

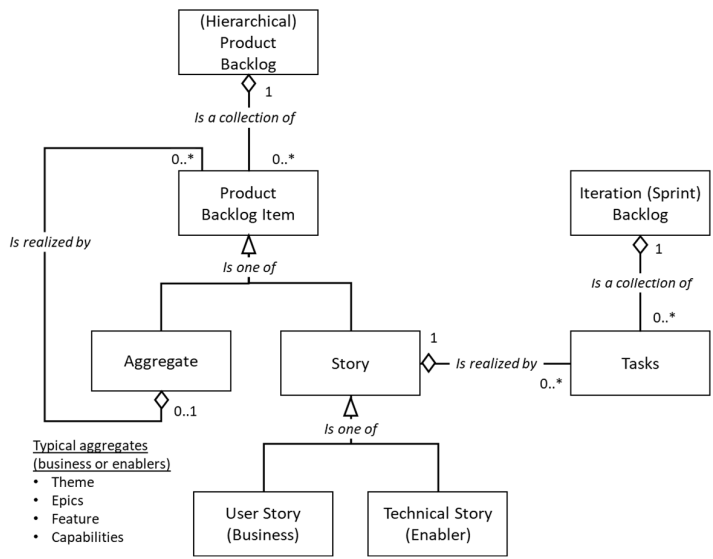

**Fig. 4.** A hierarchical product backlog and its relation to the iteration or sprint backlog

In step 6, the team will black out known non-working periods such as the holidays, training, and mandatory vacations, since in principle, there would be no work carried out during that time.

In step 7, the team will mark all the hard milestones in the work packages scheduling canvas. Hard milestones will act as anchor points for the plan.

In steps 8, 9 and 10, the team iteratively builds the Work Packages Schedule, see Figure 6, by posting sticky notes on an empty space in the work package scheduling canvas, starting first with the work corresponding to the process ceremonies, following with the work corresponding to the hard milestones and finally with that of the soft milestones. Since the planning involves the physical positioning of sticky notes on the canvas, there must be a correspondence between the work hours represented by each note and the canvas' physical dimensions. If for example, we choose a 3"x 3" sticky note to represent 40 hours of work, each three inches on the time axis of the canvas will correspond to a week and three inches in the resources axis will correspond to a full time equivalent (FTE) resource. Had we chosen a lower granularity, e.g., a sticky note to represent 150 hours of work, which would be useful in the case of a larger project, each three inches on the time axis would correspond to a month instead of a week. One could rip off sticky notes to express fractions of effort or time, but this should be hardly necessary given the resolution of the plan.

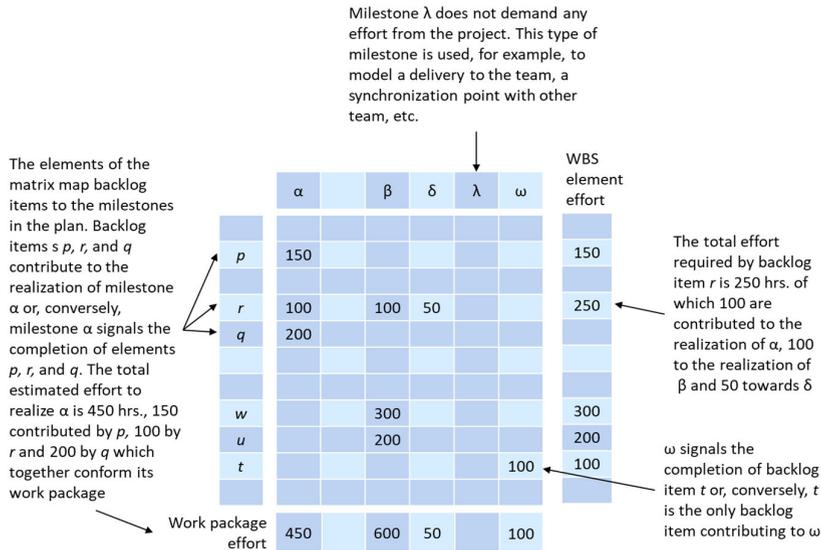

**Fig. 5.** Milestone Planning Matrix

In step 11, the plan is completed by sprucing the milestone sequence diagram, assigning due dates to the soft milestones, adding responsibility information, and integrating all of them in a common document. The approximate due date for each soft milestone is found by looking in the Work Packages Schedule for the date aligned with the right edge of the time box associated with the milestone. Hard milestones have, by definition, a set date.

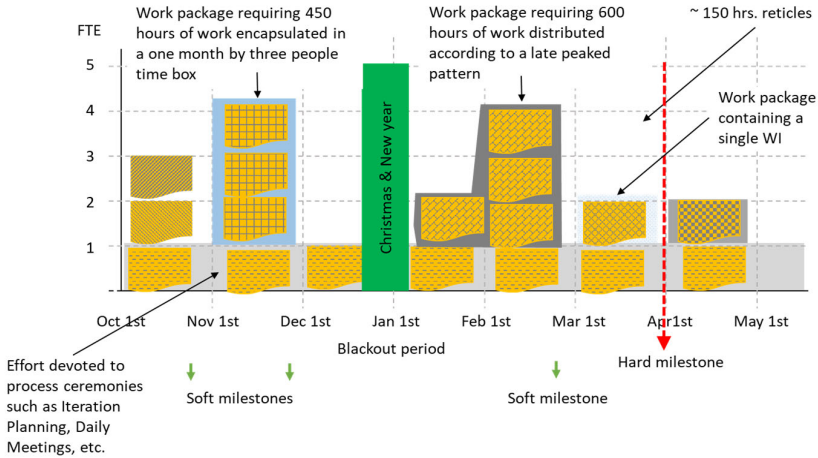

**Fig. 6.** Iterative construction of the Work Packages Schedule using sticky notes to represent the effort required to realize each milestone

## 5   Detailed example

This section provides a step-by-step account of the application of the process in figure 3 to the creation of the milestone plan in figure 1.

Imagine your company is bidding in a contract to develop an ecommerce site for a small book publisher and after some discussion with your customer you sketched the following notes:

1. Customer wants to include a beta testing period to validate the site design.
2. Customer will not accept deployment until a system wide acceptance test is satisfactorily completed.
3. Customer sign-off will follow satisfactory deployment of the system.
4. Customer would like to have at least three software releases: one to collect users' feedback via beta testing, another one to confirm the progress of the system towards the launch date and the final one to complete the system with minimum risk to the launch date.
5. In the first release she would like to include the following user stories: Category List (CL), Book Details (BD), Add to Cart (AC), Check-Out (CO) and the Data Base epic
6. In the second release, the Category Book List (CBL), Remove from Cart (RC) and Shipping Method (SM)
7. The content of the third release is not known at this time, but the customer would like to reserve a significant number of hours to incorporate the feedback resulting from beta testing and introduce other user stories that might emerge during development
8. The customer is preparing to launch its business around the beginning of May of next year so she would like the system to be ready at least one month before that.

For its part, your company:

9. Cannot start the project until the end of September and has only four developers to work on the project.
10. To minimize the risk of rework, it does not plan to start programming until the infrastructure is selected and the user interface and information architecture design are well underway.

## 5.1 Step 1

Based on the requirements above and its professional knowledge the development team created the backlog shown in Figure 7 describing their understanding of the project's scope of work and the estimated effort required for its execution. The required software capabilities were grouped into four epics: Browsing, Buying and Data Base to increase the readability of the backlog as the team felt a grouping based on Releases would have hindered the comprehension of system functionality. UX design, Infrastructure Selection, Beta and Acceptance testing are technical stories that could have been made part of the Website Software deliverable, but which the team chose to put at the first level of the backlog to highlight its understanding of the customer wishes. After some discussion about how many hours to reserve for the final release, the team decided to allocate 450 hours since this was compatible with the team's resource availability, the customer schedule and provided ample time to incorporate feedback and complete a few, yet undiscovered, features.

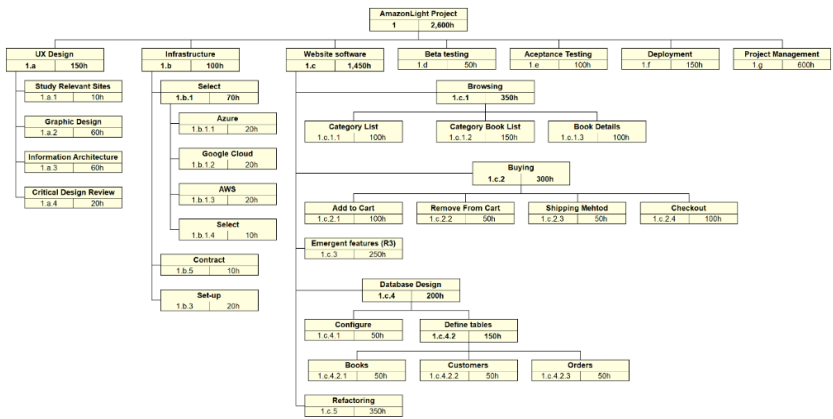

**Fig. 7.** Hierarchical backlog for the AmazonLight project

## 5.2 Step 2

After obtaining concurrence for the backlog from the sponsor, the team started choosing relevant milestones producing the list in Table 1. Beware the solution is not unequivocal. While there are self-evident milestones like project kick-off, software releases and the customer request for a beta test, others are created by the team based on its best judgment as to what is important and what is not. The completion criteria associated with each milestone defines its meaning and helps identify which backlog items should be mapped onto them. The milestones are organized in what seems like the most logical sequence to make the list easier to understand and search for.

## 5.3 Step 3

The team constructs the Milestone Dependency Diagram by organizing the milestones identified in the previous step in what seems like the most logical sequence, and then by connecting them using finish-to-finish dependencies. Notice that the dependency diagram says nothing about when the work for it ought to start. We will address this in step 10.

## 5.4 Step 4

In this step, the team reads the Milestone Dependency Diagram in its most natural sounding order, writing the milestones names from left to right as headers of the MPM. If two milestones have a similar due date, it doesn't matter which one is listed first, since the sole purpose of the ordering is to increase the matrix' readability.

## 5.5 Step 5

During this step, the team associates the backlog items with their corresponding milestones. This is a rather mechanical activity whose value resides on the visibility it brings to the planning process. As discussed above, a backlog item can be associated with more than one milestone as shown by rows 1.a, 1.d, 1.e and 1.c.5 in Figure 8. While this technique relieves us from creating backlog items just for the sake of having each item map to a single milestone, abusing it obscures the mapping, for what it should be used sparingly. Milestone "Cloud Infrastructure Available" has no backlog item associated with it, because although the work to select the infrastructure is part of the scope of the project, the effort to provision it, is not. The reason to include it as a milestone is that it represents a commitment made to the project team by the customer so it could beta test the system and to signal that a delay in fulfilling this promise could affect its completion date. In the case of project management, item 1.g, the effort was not allocated to any milestone to be later spread over the life-span of the project according to a uniform profile.

12**Table 1.** Potential milestones for the AmazonLight project

| Milestone | Rationale (from notes above) | Hard date | Completion criteria |
|---|---|---|---|
| Project kick-off | 9 | October this year | Development team assembled, meeting with project sponsor concluded |
| Design concept approved | 10 | | Information architecture and UI designed and approved by sponsor |
| Infrastructure selected | 10 | | Cloud provider selected. Consider AWS, Azure, Google Cloud and 2 others |
| Design completed | Proposed by team | | Feedback from sponsor incorporated into design |
| Cloud infrastructure available | Proposed by team | | Cloud production environment available |
| Release 1: CL, BD, AC, CO, Data Base | 4, 5 | | Indicated functionality is ready and tested at 90% coverage and working in production configuration. No broken menus or links |
| Beta testing launched | 1 | | Release 1 software made available to beta users. User behavior hypotheses defined. Website instrumentation working |
| Release 2: CBL, RC, SM | 4, 6 | | Indicated functionality is ready and tested at 90% coverage and working in production configuration |
| Beta testing results reviewed | 1 | | All insights arising from the beta testing disposed |
| Release 3: Emergent features | 7 | | Reserved effort to implement changes suggested by the beta testing and unknown features |
| Acceptance testing procedure approved | 2 | | Acceptance test suite approved by sponsor. Includes at least one positive, one negative and one invalid test case for each functionality |
| Acceptance test completed | 2 | | All acceptance test passed with no objection from sponsor |
| System deployed | 8 | April next year | All functionality running in production environment, operators trained. System must run for at least 15 consecutive days without a fault attributable to software |
| Customer sign-off | 3 | | Customer accepts ownership of the software |
| Project closed | 8 | May next year | Project postmortem executed; all records archived |

## 5.6   Step 6

The team marks the Christmas and New Year weeks as a non-working period.

## 5.7 Step 7

Beside its start and end, the project contains only one hard milestone, the "System Deployed" milestone set for the beginning of April. As by definition, a successful plan must satisfy its hard milestones, we start by marking them on the Work Package Schedule canvas as they would constrain how the work could be organized.

## 5.8 Steps 8, 9 & 10

The goal of these steps is to establish a window of opportunity in which the work stated by each work package must be executed. To do this, the team will label or somehow tag as many sticky notes as needed to cover the effort required by each work package and accommodate them in an appropriate empty space on the Work Package Schedule canvas. The tagging is important to identify which work corresponds to what, so that is possible to rearrange it if necessary. The team starts by accommodating the effort corresponding to all cross-cutting work packages, then that corresponding to the work packages connected to hard milestones and finally those corresponding to the rest of the work packages in the order dictated by the milestones' dependencies. If necessary, the team might intersperse buffers to protect milestones deemed critical.

Figure 2 shows a stylized version of a possible plan for the AmazonLight project. Notice that due to the holiday period, extending from late December to early January the effort for the Release 1 work package was spread over two months by splitting the sticky notes.

As shown by the figure, with the constraints put on the available resources – three developers and a project manager – it was not possible to deploy the system by April as the customer wanted, so after negotiations it was decided to move the milestone to the beginning of May. Other alternatives could have been to incorporate more resources, reorganize the work, e.g. relax the condition of not doing development work before the design concept has been approved, or renegotiate the scope.

## 5.9 Step 11

In Step 11, the milestone plan is completed by reading from the Work Package Schedule the approximate date in which the work associated with each milestone will be completed and assigning it as the due date for the milestone.

## 6 Summary

In this paper, we explained how the VMP method introduced in [7] can be used in an agile context. The method is based on the manipulation of reified constructs such as work packages and milestones in a collective practice, which promotes shared understanding and buy-in into the plan. The VMP method has evolved over three years of classroom and consulting experience and has been put into practice in mid-sized capstone projects, 2,500 to 5,000 person-hours long, and at two industrial organizations. It has also been evaluated for usability using the Process and Practice Usability Model [15].

Although further experience and assessments are required, its initial evaluation and observations point in the direction of the method's ease of use and its value in organizing a project.

| WBS Code | AmazonLight Project | Total Effort | Project kick-off | Design concept approved | Infrastructure selected | Cloud infrastructure available | Design completed | Release 1: CL, BD, AC, CO, Data Base | Beta testing launched | Release 2: CBL, RC, SM | Beta testing results reviewed | Release 3 | ATP approved | ATP completed | System deployed | Customer sign-off | Project closed |
|---|---|---|---|---|---|---|---|---|---|---|---|---|---|---|---|---|---|
| 1.a | UX Design | 150 | | 140 | | | 10 | | | | | | | | | | |
| 1.b | Infrastructure | 100 | | | 100 | | | | | | | | | | | | |
| 1.c.1.1 | Category List | 100 | | | | | | 100 | | | | | | | | | |
| 1.c.1.2 | Category Book List | 150 | | | | | | | | | 150 | | | | | | |
| 1.c.1.3 | Book Details | 100 | | | | | | 100 | | | | | | | | | |
| 1.c.2.1 | Add to Cart | 100 | | | | | | 100 | | | | | | | | | |
| 1.c.2.2 | Remove From Cart | 50 | | | | | | | | | 50 | | | | | | |
| 1.c.2.3 | Shipping Mehtod | 50 | | | | | | | | | 50 | | | | | | |
| 1.c.2.4 | Checkout | 100 | | | | | | 100 | | | | | | | | | |
| 1.c.3 | Emergent features (R3) | 250 | | | | | | | | | | | 250 | | | | |
| 1.c.4 | Database Design | 200 | | | | | | 200 | | | | | | | | | |
| 1.d | Beta testing | 50 | | | | | | | | 40 | | 10 | | | | | |
| 1.e | Aceptance Testing | 100 | | | | | | | | | | | | 50 | 50 | | |
| 1.f | Deployment | 150 | | | | | | | | | | | | | | 150 | |
| 1.c.5 | Refactoring | 350 | | | | | | | 150 | | 100 | | 100 | | | | |
| 1.g | Project Management | 600 | | | | | | | | | | | | | | | |
| | Work Packages | 2600 | 0 | 140 | 100 | 0 | 10 | 750 | 40 | 350 | 10 | 350 | 50 | 50 | 150 | 0 | 0 |

**Fig. 8.** Milestone Planning Matrix for the AmazonLight project